\documentclass{amsart}
\usepackage{amssymb,epsf}
\newtheorem{lemma}{Lemma}[section]
\newtheorem{theorem}[lemma]{Theorem}
\newtheorem{corol}[lemma]{Corollary}
\newtheorem{claim}[lemma]{Claim}

\newcommand{\CL}{\hbox{{$\mathcal L$}}}
\newcommand{\CI}{\hbox{{$\mathcal I$}}}
\newcommand{\CF}{\hbox{{$\mathcal F$}}}
\newcommand{\CC}{\hbox{{$\mathcal C$}}}
\newcommand{\CZ}{\hbox{{$\mathcal Z$}}}
\newcommand{\CS}{\hbox{{$\mathcal S$}}}
\newcommand{\R}{\mathbb{R}}
\newcommand{\Z}{\mathbb{Z}}
\newcommand{\C}{\mathbb{C}}
\newcommand{\cg}{\hbox{{$\mathfrak g$}}}
\newcommand{\note}[1]{}
\newcommand{\extd}{{\rm d}}
\newcommand{\image}{{\rm im\, }}
\newcommand{\isom}{{\cong}}
\newcommand{\eps}{{\epsilon}}
\newcommand{\tens}{\mathop{\otimes}}
\newcommand{\id}{{\rm id}}
\newcommand{\<}{\langle}
\renewcommand{\>}{\rangle}
\newcommand{\del}{\partial}
\newcommand{\End}{{\rm End}}
\newcommand{\Dsl}{{D\kern-6pt/}}
\newcommand{\cliff}{{\rm Cliff}}

\renewcommand{\o}{{}_{\scriptscriptstyle(1)}}
\renewcommand{\t}{{}_{\scriptscriptstyle(2)}}
\renewcommand{\th}{{}_{\scriptscriptstyle(3)}}
\newcommand{\bo}{{}^{\bar{\scriptscriptstyle(1)}}}
\newcommand{\bt}{{}^{\bar{\scriptscriptstyle(\infty)}}}
\newcommand{\und}[1]{{\underline {#1}}}
\newcommand{\rcross}{{\triangleright\!\!\!<}}

\newcommand{\cobicross}{{\triangleright\!\!\!\blacktriangleleft}}

\newcommand{\eqn}[2]{\begin{equation}#2\label{#1}\end{equation}}

\begin{document}

\title[NONCOMMUTATIVE PHYSICS ON $(\Z_2)^n$ LATTICES AND
CLIFFORD ALGEBRAS]{\rm \large NONCOMMUTATIVE PHYSICS ON LIE
ALGEBRAS, $(\Z_2)^n$ LATTICES AND CLIFFORD ALGEBRAS}
\author{S. Majid}
\address{School of Mathematical Sciences\\
Queen Mary, University of London\\ 327 Mile End Rd,  London E1
4NS, UK.}

\thanks{This paper is in final form and no version of it will be submitted for publication
elsewhere} 

\date{2/2003}%

\maketitle

\begin{abstract}
We survey noncommutative spacetimes with coordinates being
enveloping algebras of Lie algebras. We also explain how to do
differential geometry on noncommutative spaces that are obtained
from commutative ones via a Moyal-product type cocycle twist, such
as the noncommutative torus, $\theta$-spaces and Clifford
algebras. The latter are noncommutative deformations of the finite
lattice $(\Z_2)^n$ and we compute their noncommutative de Rham
cohomology and moduli of solutions of Maxwell's equations. We
exactly quantize noncommutative $U(1)$-Yang-Mills theory on
$\Z_2\times\Z_2$ in a path integral approach.
\end{abstract}

\section{Introduction}

Noncommutative geometry has made rapid progress in the past two
decades and is arguably now mature enough to be fully computable
and relevant to real physical experiments. The benefit of
noncommutative geometry is that whereas the usual coordinates on a
space are commutative and most `nice' commutative algebras could
be viewed that way, noncommutative geometry, by relaxing the
commutativity assumption, allows practically any algebra to be
viewed geometrically. This opens up a brave new world of
possibilities for model-building in which our favorite
noncommutative algebras, such as matrices, angular momentum
operators, Clifford algebras, can be viewed as noncommutative
coordinates with all of the geometry that that entails. And why
should we need such a possibility? First of all, we already
encounter such noncommutative geometries when we quantise a
system. Recall that classical mechanics is described by classical
symplectic geometry, Hamilton-Jacobi equations of motion, etc.
What happens to all this geometry when we quantise, is it all just
thrown away after we have got the right commutators? The
corrspondence principle in quantum mechanics says that it should
not be thrown away, that classical variables such as angular
momentum should have their analogues in the quantum system with
analogous properties. But this is a rather vague statement.
Analogues of what? It seems likely that the quantum system should
be as rich if not richer in structure than the classical one and
should have among other things analogues of all of the geometry of
phase space, Hamiltonian flows etc. if only we had the `eyes' to
see it. Noncommutative geometry can in principle provide those
eyes.

In this article we will not apply noncommutative geometry to
quantum phase space exactly, though that is an interesting
direction that deserves more study. Instead we will look at the
even more radical proposal that spacetime itself may have
noncommutative coordinates, an as yet undiscovered but potentially
new physical phenomenon if it were ever verified. It could be
called `cogravity' for reasons explained in Section~2. The idea is
not new \cite{Ma:ista}, but we shall focus on three simple
examples based on Lie algebras viewed as noncommutative
spacetimes, with some new observations for the $\theta$-spacetime
case.

After the preliminary Section~2, we shall move to new results. The
first, in Section~3 is an application of noncommutative geometry
to quantize $U(1)$-Yang-Mills theory on the finite set $\Z_2\times
\Z_2$ all the way up to Wilson loop vacuum expectation values.
Section~4 covers the electromagnetic theory on Clifford algebras
$\cliff(n)$ viewed as noncommutative coordinates. This is not only
for fun; Clifford algebras can be viewed as `Moyal-product'
cocycle type quantizations of $\Z_2^n$ in the same spirit as
recently popular proposals for $\theta$-spacetime in string
theory, but now in a discrete version. Section~5 is a more
mathematical section for experts and is the general theory behind
such cocycle twist quantisations. This theory depends on
Drinfeld's quantum groups work but its full impact remains not so
well-known in the operator theory approach to noncommutative
geometry of Connes \cite{Con} and others. We point out that
cocycle twisting means by the Majid-Oeckl twisting
theorem\cite{MaOec:twi} that the differential geometry also twists
(or gets `quantised') by the same cocycle and that this covers the
algebraic (but not functional analytic) aspects of many famous
examples including the celebrated noncommutative torus $A_\theta$.

Next, a few words about the relation between the operator theory
and quantum groups approaches to noncommutative geometry. From a
mathematical point of view the former began in the 1940's with the
theorem of Gelfand and Naimark characterising functions on locally
compact spaces as commutative $C^*$ algebras and hence proposing
any noncommutative $C^*$ algebra as a `noncommutative space'.
Vector bundles were characterised in the 1960's by the Serre-Swann
theorem as finitely generated projective modules, a notion again
working in the noncommutative case. Operator K-theory led in the
1980's to cyclic cohomology and ultimately to Connes formalisation
of the Dirac operator on a spin manifold in operator algebra terms
as a `spectral triple'\cite{Con}. At the same time in the 1980's
there emerged from the theory of integrable systems a large class
of examples of noncommutative algebras with manifest `geometrical'
content, namely quantum groups $\C_q[G]$ deforming the usual
coordinate ring of a simple Lie group $G$. Also at the same time
in the 1980s there emerged a different `bicrossproduct' class of
quantum groups \cite{Ma:pla} again built on Lie groups but this
time from a self-duality approach to quantum gravity and
Planck-scale physics. Moreover, just as Lie groups and their
homogeneous spaces provide key examples of classical differential
geometry, so quantum groups {\em should} provide key examples of
noncommutative geometry. This `quantum groups approach' to
noncommutative geometry starts with differential structures on
quantum groups \cite{Wor:dif}, gauge theory with quantum group
fiber\cite{BrzMa:gau} and eventually a notion of a `quantum
manifold' as an algebra equipped with a quantum group frame bundle
\cite{Ma:rief}. It is important for us here, however, that whereas
the $q$-deformation literature and examples like $\C_q[SU_2]$
played an important role in developing the right axioms, the final
theory, like the operator algebras approach, applies in principle
to any algebra including ones that have nothing whatever to do
with $q$-deformations. It should be said also that the operator
theory approach tends to be mathematically more sophisticated
while the quantum groups one tends to be more computational and
examples-led. One of the aims of Section~5 is to promote
convergence between the two approaches.

We will need some precise notations. Let $M$ be a unital algebra
over $\C$ (say). We regard $M$ as `coordinate algebra' on a space,
even though it need not be commutative. By `differential calculus'
we mean $(\Omega^1,\extd)$ where $\Omega^1$ is an $M-M$ bimodule,
$\extd:M\to \Omega^1$ obeys the Leibniz rule \eqn{leib}
{\extd(fg)=(\extd f)g+f\extd g,\quad \forall f,g\in M} and
elements of the form $f\extd g$ span $\Omega^1$. We also want in
practice that the kernel of $\extd$ is spanned by $1$, the
constants. Note that we only need $\Omega^1$ to be a bimodule --
indeed, assuming $f\extd g=(\extd g)f$ for all $f,g$ would imply
$\extd(fg-gf)=0$ which would make $\extd$ trivial on a very
noncommutative algebra. So we do not assume this, but only the
weaker bimodule condition $f((\extd g)h)=(f\extd g)h$ for all
$f,g,h$. This is the starting point of all approaches to
noncommutative geometry. After this, one can extend to an entire
exterior algebra of differential forms $\Omega,\extd$ with $\extd$
a super-derivation of degree 1 with $\extd^2=0$. How this is done
is different in the two approaches; in the quantum groups one we
use considerations of a background symmetry to narrow down the
possible extensions with the result that in most cases there is a
reasonably natural choice. In the Connes approach it is defined by
choosing an operator to be the `Dirac' operator. Another
ingredient one needs is a Hodge $\star$ operator $\Omega^m\to
\Omega^{n-m}$ where $n$ is the top degree (which we assume exists,
and which is called the volume-dimension of the noncommutative
space).

With these few ingredients one can already do a lot of physics, in
particular most of electromagnetism. The simplest version of this
which we call `Maxwell theory' is to think of a gauge field $A\in
\Omega^1$ modulo exact forms, with curvature $F=\extd A$. Next
there is a nonlinear version which we call $U(1)$-Yang-Mills
theory where we consider $A\in\Omega^1$ modulo the transformation
\eqn{Agauge}{A\mapsto uAu^{-1}+u\extd u^{-1}} with $u$ an
invertible element of $M$. Here the curvature is $F=\extd
A+A\wedge A$ and transforms by conjugation. Such a theory has all
the flavour of a nonAbelian gauge theory but not because of a
nonAbelian gauge group, rather because differentials and functions
do not commute. One can go on and do nonAbelian gauge theory,
Riemannian geometry and construct a Dirac operator $\Dsl$. This
would be beyond our present scope but we refer to concrete models
such as in \cite{Ma:rief}\cite{NML:rie}\cite{Ma:ric}. It should be
mentioned that the results do not usually fit precisely the axioms
of a spectral triple, but perhaps something like them.

\section{Noncommutative spacetime and cogravity}

This section is by way of `warm up' for the reader to get
comfortable with the methods in a more familiar setting before
applying it to other algebras. We look particularly at models of
noncommutative spacetime where the coordinate algebra is the
enveloping algebra $U(\cg)$ and $\cg$ is a Lie algebra, with
structure constants $c_{ij}{}^k$ say. Putting in a parameter
$\lambda$, it means commutation relations \eqn{liereln}{
[x_i,x_j]=\lambda\, c_{ij}{}^k x_k} This is a well-studied system
in mathematical physics, namely $U(\cg)$ is the standard
quantisation of the Kirillov-Kostant Poisson bracket on $\cg^*$
defined canonically by the Lie algebra structure. The symplectic
leaves for this are the coadjoint orbits and their quantisation is
given by setting the Casimir of $U(\cg)$ to a fixed number.

What if actual space or spacetime coordinates $x_i,t$ might be
elements of such a noncommutative algebra instead of numbers? As
in quantum mechanics it means of course that they cannot be
simultaneously diagonalised so that there will be some uncertainty
or order-of-measurement dependence in the precise location of an
event. But there would be many more effects as well, depending on
the noncommutative algebra used. We nevertheless propose:

\begin{claim} (A) Noncommutative $x_i,t$ would be a new (as yet undiscovered)
physical effect `cogravity'

 (B) Even if originating in quantum gravity
corrections of Planck scale order, the effect {\em could} in
principle be tested by experiments today.
\end{claim}

We will look at the second claim in Section~2.1 by way of our
first example, the Majid-Ruegg $\kappa$-spacetime. Here we look
briefly at claim (A). The first thing to note is that such
spacetimes $U(\cg)$ are noncommutative analogues of {\em flat}
space. This is because they are Hopf algebras with a trivial
additive coproduct and counit \eqn{envcoprod}{ \Delta 1=1\tens
1,\quad \eps 1=1,\quad \Delta \xi=\xi\tens 1+1\tens \xi,\quad
\eps\xi=0,\quad \forall\ \xi\in \cg} extended multplicatively,
which is to say an additive abelian `group' structure on the
noncommutative space. On the other hand, of these were coordinates
of spacetime then the dual Hopf algebra should be the momentum
coordinate algebra. Here it is the usual commutative coordinate
ring $\C[G]$ of the underlying Lie group $G$, which has a
nonAbelian multiplicative coproduct for group multiplication. So
momentum space is a nonAbelian group manifold with curvature (at
least in compact cases like $SU_2=S^3$), but an ordinary
(commutative) space. Moreover, there is a nonAbelian Fourier
transform \eqn{envFou}{ \CF:U(\cg)\to \C[G]} or more precisely on
a completion of these algebras to allow functions with good decay
properties. We see that the physical meaning of noncommutative
space coordinates is equivalent under Fourier transform to
curvature or gravity in momentum space. The reader is probably
more familiar with the flipped situation in which position space
is classical and curved, such as a nonAbelian group $SU(2)$, and
its Fourier dual is a noncommutative space with noncommuting
momenta or covariant derivatives, such as
$[p_i,p_j]=\imath\eps_{ijk}p_k$. The idea of noncommutative space
or spacetime is mathematically just the same but the with roles of
position and momentum swapped. We summarise a typical situation in
Table 1 to give some idea of the physical meaning of the
mathematics that we propose here. It makes it clear that this
reversed possibility of  noncommutative position space and curved
momentum space  is a potentially new physical effect dual to
gravity. Associated to it is a new dimensionful parameter, say
$\lambda$ controlling the noncommutativity of spacetime or
curvature of momentum space. It is independent of usual curvature
in position space, which is associated to the Newton constant
parameter, and to Planck's constant which controls
noncommutativity between the position and momentum sectors. This
point of view was explained in detail in \cite{Ma:ista}.

\begin{figure}\[\begin{array}{l|c|c}
&{\rm Position}& {\rm Momentum}\\
\hline
{\rm Gravity} & {\rm Curved}& {\rm Noncommutative} \\
& \sum_i x_i^2=1 & [p_i,p_j]=\imath\eps_{ijk}p_k \\
\hline
{\rm Cogravity}& {\rm Noncommutative}& {\rm Curved}\\
& {}[x_i,x_j]=\imath\eps_{ijk}x_k& \sum_i p_i^2=1\\
\hline {\rm Qua. Mech.}&\multicolumn{2}{c}{
[x_i,p_j]=\imath\hbar\delta_{ij}}
\end{array}\]
\caption{Gravity and cogravity related by nonAbelian Fourier
transform}
\end{figure}

Turning to technical details, we use the following  general
construction which comes out of the analysis of
translation-invariant differential structures on quantum groups.
If $M=U(\cg)$, a differential calculus is specified by a right
ideal $\CI\subset U(\cg)^+$ where $U(\cg)^+$ denotes expressions
in the generators with no constant term. Let
$\Lambda^1=U(\cg)^+/\CI$ be the quotient and $\pi:U(\cg)^+\to
\Lambda^1$ be the projection map. Let $\tilde\pi$ be the
projection from $U(\cg)$ where we first apply $\id-1\eps$ which
projects to $U(\cg)^+$, and then $\pi$. We have \eqn{envcalc}{
\Omega^1=U(\cg)\Lambda^1,\quad \extd f=(\id\tens \tilde\pi)\Delta
f,\quad \omega f=f\o\pi(\tilde\omega.f\t),\quad\forall f\in
U(\cg),\ \omega\in\Lambda^1,} where $\tilde\omega$ is a
representative projecting onto $\omega$ and $\Delta f=f\o\tens
f\t$ is a notation. The elements of $\Lambda^1$ are the `basic
1-forms' and others are given by these with `functional'
coefficients from $U(\cg)$ on the left. Note that 1-forms and such
`functions' do not commute. The wedge product for the entire
exterior algebra is given in the present class of examples simply
by elements of $\Lambda^1$ anticommuting among themselves. The
Hodge $\star$ operator is similarly the usual one among such basic
one-forms. Finally, given a basis $\{e_\mu\}$ of $\Lambda^1$, we
define partial derivatives $\del^\mu$ by (summation understood):
\eqn{partial}{ \extd f= (\del^\mu f)e_\mu,\quad \forall f\in
U(\cg).}

A natural way to specify the ideal $\CI$ is as the kernel of a
representation $\rho:\cg\to \End(V)$ extended as a representation
of $U(\cg)^+$. Then we can identify $\Lambda^1$ as the image of
$\rho$ and have the formulae \eqn{liecalc}{
\extd\xi=\rho(\xi),\quad [\omega,\xi]=\omega.\rho(\xi),\quad
\forall \xi\in\cg,\ \omega\in \image(\rho),} where  we take the
matrix product on the right.

\subsection{$\lambda$-spacetime and gamma-ray bursts}

We start with one of the most accessible noncommutative spacetime
in this family\cite{MaRue:bic} \eqn{lmink}{ [t,x_i]=\imath\lambda
x_i,\quad [x_i,x_j]=0} where $\lambda$ has time dimension. If the
effect is generated by quantum gravity corrections, we might
expect $\lambda\sim 10^{-44}s$, the Planck time. One can also work
with $\kappa=\lambda^{-1}$. Such a spacetime in two dimensions
(i.e. the enveloping algebra of $U(b_+)$ where $b_+\subset su_2$)
was first proposed in \cite{Ma:reg} with an additional $q$
parameter which one may set to 1. The first thing the reader will
be concerned about is that this proposal manifestly breaks Lorentz
invariance, so cannot be correct. What was shown in
\cite{MaRue:bic} that there {\em is} not only Lorentz but a full
Poincar\'e invariance, but under a quantum group
\eqn{kpoinc}{U(so_{1,3})\cobicross \C[\R^{1,3}].} This was shown
also to be isomorphic to a `$\kappa$-Poincar\'e' quantum group
that had been proposed from another point of view (that of
contraction from $U_q(so_{2,3})$) by J. Lukierski et al. in
\cite{LNRT:def} but without a noncommutative spacetime on which to
act.

What is important is that all of the geometrical consequences are
not ad hoc but naturally follow within our approach to
noncommutative geometry from the choice (\ref{lmink}) of algebra.
The first step it to compute the natural differential structure.
There is not much choice and we take the 4-dimensional
representation \eqn{larho}{ \rho(x_\mu)=\imath\lambda\begin{pmatrix}0\\
e_\mu
\end{pmatrix}} where the $e_\mu=(0\cdots1\cdots 0)$ has 1 is in
the $\mu+1$-th position and $\mu=0,1,2,3$. We write $x_0\equiv t$.
The basic 1-forms are provided by the $\extd x_\mu$ as certain
matrices, and these span the image of $\rho$ since the
exponentiation of the Lie algebra has the similar form in this
representation. The result from the general theory above is then
\eqn{lacalc}{ (\extd x_j)x_\mu=x_\mu\extd x_j,\quad (\extd
t)x_\mu-x_\mu \extd t=\imath\lambda \extd x_\mu.} The form for
$\extd$ then implies \eqn{lapartial}{
\del^i:f(x,t):=:{\del\over\del x_i}f(x,t):,\quad \quad
\del^0:f(x,t):=:{f(x,t+\imath\lambda)-f(x,t)\over\imath \lambda}:}
for normal ordered polynomial functions. We use such normal
ordered functions, with $t$ to the right, to describe a general
function in the spacetime. Under this identification we can extend
all formulae to formal power-series. Note that we see the effect
of the noncommutative spacetime as forcing a lattice-like finite
difference for the time derivative, and that this is actually by
an imaginary time displacement. This is similar to to the
$+\imath\eps$ prescription in quantum field theory where
operations more naturally take place in Euclidean space and must
be Wick rotated back to the Minkowski picture. Note also that the
noncommutativity of functions and the time direction generates the
exterior derivative in the sense \eqn{lainner}{ [\extd
t,f]=\imath\lambda \extd f,\quad \forall f} which is a typical
feature of many noncommutative geometries but has no classical
analogue.

Next, at least formally, we have eigenfunctions of the $\del^\mu$
given by \eqn{laplane}{ \psi_{k,\omega}=e^{\imath k x}e^{\imath
\omega t},\quad
\psi_{k,\omega}\psi_{k',\omega'}=\psi_{k+e^{-\lambda\omega}k',\omega+\omega'},\quad
(\psi_{k,\omega})^{-1}=\psi_{-ke^{\lambda\omega},-\omega}.} where
we also show the product and inversion of such functions. We see
from the latter that the Fourier dual or momentum space is the
nonAbelian Lie group $\R\rcross_\lambda \R^3$. The invariant
integration is the usual one on normal ordered functions and
hence, allowing for the required ordering, the Fourier transform
is given by \begin{eqnarray}\label{laFou}
\CF(:f(x,t):)(k,\omega)&=&\int \psi_{k,\omega}:f(x,t):=\int \extd
x\extd t\,  e^{\imath k\cdot x}e^{\imath \omega
t}f(e^{-\lambda\omega}x,t)\nonumber \\
&=&e^{\lambda\omega}\CF_{\rm
usual}(f)(e^{\lambda\omega}k,\omega),\end{eqnarray} i.e. reduces
to a usual Fourier transform. We also compute the scalar wave
operator from $\star\extd\star\extd$ and obtain the usual form
$(\del^0)^2-\sum_i (\del^i)^2$, which now has massless modes given
by plane waves with \eqn{lawave}{
{2\over\lambda^2}(\cosh(\lambda\omega)-1)-k\cdot k
e^{\lambda\omega}=0.} This is a straightforward application of the
Fourier theory on nonAbelian enveloping algebras introduced in
\cite{Ma:ista}.

More details and, in particular, a physical analysis, appeared in
\cite{AmeMa:wav}. Critically, one has to make a postulate for how
the mathematics shall be related to experimental numbers. Here,
given the solvable Lie algebra structure, we proposed that
expressions shall be identified only when normal ordered. In
effect, one measures $t$ first in any experiment. Under such an
assumption one can analyse the wave-velocity of the above plane
waves and argue that the dispersion relation has the classical
form. Both of these steps are needed for any meaning to
predictions from the theory. We can then find for the massless
wave speed: \eqn{lightspeed}{ |{\extd \omega\over \extd
k}|=e^{-\lambda\omega}} in units where 1 is the usual speed of
light. We assumed that light propagation has the same features as
our analysis for massless fields, in which case the physical
prediction is that the speed of light depends on energy.

One may then, for example, plug in numbers from gamma-ray burst
data as follows. These gamma-ray bursts have been shown in some
cases to travel cosmological distances before arriving on Earth,
and have a spread of frequencies from 0.1-100 MeV in energy terms.
According to the above, the relative time delay $\Delta_t$ on
travelling distance $L$ for frequencies $\omega$,
$\omega+\Delta_\omega$ is \eqn{gammaburst}{ \Delta_t\sim\lambda
\Delta_\omega {L\over c}\sim 10^{-44}{\rm s} \times 100 {\rm
MeV}\times  10^{10}{\rm y}\sim 1\ {\rm ms}} where we put in the
worst case for $\lambda$, namely the Planck time. We see that
arrival times would be spread by the order of milliseconds, which
is in principle observable! To observe it would need a statistical
analysis of many gamma-ray burst events, to look for an effect
that was proportional to distance travelled (since little is known
about the initial creation profile of any one burst). This in turn
would require accumulation of distance-data for each event by
astronomers, such as has been achieved in some cases by
coordination between the (now lost) BEPPO-SAX satellite to detect
the gamma-ray burst and the Hubble telescope to lock in on the
host galaxy during the afterglow period. With the design and
implementation of such experiments and statistical analysis, we
see that one might in principle observe the effect even if it
originates in quantum gravity.

Let us mention finally that there are many other effects of
noncommutative spacetime, some of which might be measured in
earthbound experiments. For example, the LIGO/VIRGO gravitational
wave interferometer project, although  intended to detect
gravitational waves, could also detect the above variable speed of
light effect; a detailed theoretical model has yet to be built,
but some initial speculations are in \cite{Ame:nat}. Similarly,
reversal of momentum in our theory is done by group inversion,
which means $(k,\omega)\to (-ke^{\lambda\omega},-\omega)$, a
modification  perhaps detectable as CPT violation in neutral Kaon
resonances\cite{AmeMa:wav}. The problem of interpretation in
scattering theory is still open, however: what is the meaning of
nonAbelian momentum and how might one detect it? Let us not forget
also that the usual Lorentz and Poincar\'e group covariance is
modified to a certain quantum group. It contains the usual
$U(so_{1,3})$ as a sub-Hopf algebra but acting in a modified
non-linear way, which means that special relativity effects are
slightly modified. This is another source of potential
observability. In short, we have indicated the reasons for Claim
(B) above, but much needs to be done by way of physical
interpretation and experimental design.

Finally, we comment on the rest of the geometry. The exterior
algebra and cohomology holds no surprises (the latter is trivial).
Indeed, the space is geometrically as trivial as $\R^{1,3}$. This
is consistent with our view that the above predictions have
nothing to do with gravity, it is an independent effect.  Thus,
the curvature in the Maxwell theory of a gauge field $A=A^\mu
\extd x_\mu$ is \eqn{kappaF}{ F=\extd A=\del^\mu A^\nu \extd
x_\mu\wedge \extd x_\nu} and its components have the usual
antisymmetric form because the basic forms anticommute as usual.
Because the Hodge $\star$ operations on them are also as usual
when we keep all differentials to the right, and because the
partial derivatives  commute, the Maxwell operator
$\star\extd\star\extd $ on 1-forms has the same form as the usual
one, namely the scalar wave operator as above if we take $A$ in
Lorentz gauge $\del^\mu A_\mu=0$. This is why Maxwell light
propagation is as in the scalar field case as assumed above.  If
we take a static electric source $J=\rho(x)\extd t$ then the
scalar potential and electric flux are as usual, since the spatial
derivatives are as usual. Magnetostatic solutions likewise have
the same form. Mixed equations with time dependence have the usual
form but with $\del^0$ for the time derivative. The
$U(1)$-Yang-Mills theory appears more complicated but has similar
features to the Maxwell one. Now the curvature is \eqn{kappaYM}{
F=(1+\imath\lambda A^0)\extd A+ [A^0,A^i] \extd t\wedge \extd x_i
+{1\over 2}[A^i,A^j]\extd x_i\wedge \extd x_j} where the extra
terms are from $A\wedge A$ using the relations (\ref{lainner})
between functions and 1-forms. A gauge transformation is
\eqn{kappaA}{ A^i\mapsto uA^i u^{-1}+u(1+\imath\lambda A^0)\del^i
u^{-1},\quad A^0\mapsto uA^0u^{-1} +u(1+\imath\lambda
A^0)\del^0u^{-1}.} The Dirac operator and spinor theory requires
more machinery and has not yet been worked out in any meaningful
(not ad-hoc) manner. Likewise, quantum field theory on the
noncommutative Minkowski space is possible, starting with the
Fourier transform above, but has not been fully worked out.

\subsection{Angular momentum space and fuzzy spheres}

Next we look at angular momentum operators but now regarded in a
reversed role as noncommutative position space, which means the
Lie algebra $su_2$ with relations \eqn{suspace}{
[x_i,x_j]=\imath\lambda \eps_{ijk}x_k.} One may add a commutative
time coordinate if desired, but the first remarkable discovery is
that this is not required: when one makes the analysis of
differential calculi there is only one natural choice and it is
already four, not three dimensional! The algebra itself needs no
introduction, but some aspects have been studied under the heading
`fuzzy spheres'. More precisely, these are finite-dimensional
matrix algebras viewed as the image of (\ref{suspace}) in a fixed
spin representation, in which case one is seeing effectively the
quotient where the Casimir $x\cdot x$ is equal to a constant. We
are not taking this point of view here but working directly with
the infinite-dimensional coordinate algebra (\ref{suspace})
itself. This model of noncommutative geometry appeared recently in
\cite{BatMa:ang} and we give only a brief synopsis of a few
aspects. First of all, the reader may ask about the Euclidean
group invariance. This is preserved, but again as a quantum group
\eqn{Dsu2}{ U(su_2)\rcross \C[SU_2]} where $SU_2$ is a curved
momentum space (as promised above). This is an example of a
Drinfeld quantum double as well as a partially trivial
bicrossproduct. As $\lambda\to\infty$ it becomes an $S^3$ of
infinite radius, i.e. flat $\R^3$ acting by usual translations as
it should. The $U(su_2)$ acts by the adjoint action which becomes
usual rotations in the limit. We see that nontrivial quantum
groups arise in very basic physics wherever noncommutative
operators obeying the angular momentum relations are present.

For the differential geometry, we take $\rho(x_i)={\lambda\over
2}\sigma_i$ the usual Pauli-matrix representation and the basic
1-forms $\Lambda^1=M_2$, the space of $2\times 2$ matrices since
the image of $\rho$ in this case is everything. Then
\eqn{sucalc}{\extd x_i={\lambda\over 2} \sigma_i,\quad (\extd x_i)
x_j-x_j \extd x_i=\imath{\lambda\over 2} \eps_{ijk}\extd
x_k+{\lambda\over 4}\delta_{ij}e_0,\quad e_0 x_i-x_i e_0=\lambda
\extd x_i,} where $e_0$ is the $2\times 2$ identity matrix which,
together with the Pauli matrices $\sigma_i$ completes the basis of
basic 1-forms. It provides a natural time direction, even though
there is no time coordinate. Indeed, the first cohomology is
nontrivial and spanned by $e_0$, i.e. it is a closed 1-form which
is not $\extd$ of anything (it is denoted by $\theta$ in
\cite{BatMa:ang}). Nevertheless, like $\extd t$ in the previous
section, we see from (\ref{sucalc}) that $e_0$ generates the
exterior derivative by commutator, \eqn{suinner}{[e_0,f]=\lambda
\extd f,\quad \forall f\in U(su_2).}

The partial derivatives defined by \[ \extd f=(\del^i f)\extd
x_i+(\del^0 f)e_0\] for all $f$ are hard to write down explicitly;
they are given in \cite{BatMa:ang}. Nevertheless, the formal group
elements \eqn{suplane}{ \psi_{k}=e^{\imath k.x}} are the plane
waves and eigenfunctions for the partial derivatives. For the same
reasons as in our previous model in Section~2.1, the scaler
Laplacian comes out as $(\del^0)^2-\sum_i (\del^i)^2$ when we take
a local Minkowski metric. Its value on plane waves is
\eqn{suwave}{ {1\over\lambda^2}\left((\cos({\lambda|k|\over
2})-1)^2+4 \sin^2({\lambda |k|\over 2})\right).} The Maxwell
theory may likewise be worked out and has similarities with the
usual one due to the fourth dimension provided by $e_0$, except
that we will only have static solutions since we have no time
variable. This time the coordinates are fully `tangled up' by the
relations (\ref{suspace}) and solutions are rather hard to write
down explicitly. One solution, for a uniform electric charge
density and spherical boundary conditions at infinity (i.e.
constructed as a series of concentric shells) has timelike source
 $J=e_0$, scalar potential $x\cdot x$ and electric field
proportional to $x$, i.e. radial, see \cite{BatMa:ang}. There is
similarly a magnetic solution for a uniform current density.
Non-uniform solutions have yet to be worked out due only to their
algebraic complexity. The Dirac operator is known and given in
\cite{BatMa:ang} also, as are coherent states in which the
noncommutative coordinates behave as close as possible to
classical. The explicit form of $\del^0$ is also interesting and
takes the form \eqn{del0su2}{ \del^0 f={\lambda\over 8}\sum_i
(\del^i)^2 f+O(\lambda^2)} which is the free particle Hamiltonian
to lowest order. This is a general feature of many noncommutative
algebras, that there is an extra cotangent direction $e_0$ induced
by the noncommutative geometry as generating $\extd$ by
commutator, and one could even say that this is the `origin of
time evolution' if one defines the corresponding energy as
$\del^0$ and asks that it be $\del\over\del t$ in the algebra with
a variable $t$ adjoined. Details of such a philosophy will appear
elsewhere.

Finally, we note that there are a couple of physical models in
which this kind of noncommutative space could appear naturally.
One is $2+1$ quantum gravity in a Euclidean version based on an
$iso(3)$-Chern-Simons theory. There one finds\cite{Sch:com} that
the quantum states have a quantum group symmetry, namely the
double (\ref{Dsu2}), suggesting that our above model should
provide a description of the relevant effective geometry. The
other is with a certain form of ansatz for matrix models in string
theory, under which the theory reduces to one on a fuzzy sphere.
On the other hand, the problem of formulating the physical
consequences of the above noncommutative geometry is independent
of the underlying theory of which it may be an effective model.

\subsection{$\theta$-space and the Heisenberg algebra}

The first proposal for spacetime was probably made by
Snyder\cite{Sny} in the 1940's even before the modern machinery of
noncommutative geometry, and took the form \eqn{thspace}{
[x_\mu,x_\nu]=\imath \theta_{\mu\nu},\quad
\theta_{\mu\nu}=-\theta_{\nu\mu}\in \C} where $\theta_{\mu\nu}$
were operators with further properties arranged in such a way as
to preserve Lorentz covariance. More recently such algebras have
been revived by string theorists with $\theta$ now a number and
called noncommutative `$\theta$-space'. This is no longer any kind
of noncommutative Euclidean or Minkowski space since it does not
appear to have any (pseudo)orthogonal group or quantum group
appropriate to that. Rather, it is just the usual Heisenberg
algebra of quantum mechanics under another context and has a
symplectic character. It can also be viewed as a noncommutative
torus in an unexponentiated form. For our present treatment we
assume that the space is $2n$-dimensional and $\theta_{\mu\nu}$
nondegenerate. We take the latter in normal form and replaced by a
single central variable, say $t=x_0$, i.e. we take
\eqn{thlaspace}{ [x_i,x_j]=0,\quad [x_i,x_{-j}]=\imath\lambda
\delta_{ij} t,\quad [x_{-i},x_{-j}]=0} in terms of new variables
grouped as positive and negative index. This is now of our
enveloping algebra form generated by a $2n+1$-dimensional
Heisenberg Lie algebra. We will also give a different treatment of
(\ref{thspace}) by twisting theory in Section~5.

For the calculus we take the standard representation
\eqn{thrho}{ \rho(x_i)=\imath\lambda\begin{pmatrix}0 & e_i& 0 \\ 0&0&0\\
0&0&0\end{pmatrix},\quad
\rho(x_{-i})=\imath\lambda\begin{pmatrix}0&0&0\\ 0&0&e^t_i\\
0&0&0\end{pmatrix},\quad
\rho(t)=\imath\lambda\begin{pmatrix}0&0&1\\0&0&0 \\
0&0&0\end{pmatrix}} where $e_i$ is a row vector with 1 in the
$i$-th position and $e_i^t$ is its transpose. The general
construction then gives the basic 1-forms $\extd x_0$, $\extd
x_i$, $\extd x_{-i}$ as   certain matrices. They span the image of
$\rho$ since the Heisenberg Lie algebra exponentiates to a similar
form  in this representation. We obtain \eqn{thcalc}{(\extd
x_{-i}) x_{\pm j}=x_{\pm j}\extd x_{-i},\quad (\extd x_i) x_j=x_j
\extd x_i,\quad (\extd x_i) x_{-j}-x_{-j}\extd x_i =\imath\lambda
\delta_{ij}\extd t} and $\extd t$ central. The partial derivatives
defined by \[ \extd f=(\del^i f)\extd x_i+(\del^{-i}f)\extd
x_{-i}+(\del^0f)\extd t\] then turn out to be just the usual
derivatives on functions provided these are normal ordered with
all $x_{-i}$ to the left of all $x_j$. This is even simpler than
our first example above.

The plane wave eigenfunctions of the partial derivatives are the
group elements \eqn{thpartial}{ \psi_{k_-,k_+,\omega}=e^{\imath
k_-\cdot x_-}e^{\imath k_+\cdot x_+}e^{\imath\omega t}, \quad
\psi_{k_-,k_+,\omega}\psi_{k'_-,k'_+,\omega'}
=\psi_{k_-+k_-',k_++k_+',\omega+\omega'+\lambda k_+\cdot k_-'}}
for the Heisenberg group. Integration is the usual one on normal
ordered functions and hence, allowing for this, the Fourier
transform is given by \begin{eqnarray}
\CF(:f:)(k_-,k_+,\omega)&=&\int
\psi_{k_-,k_+,\omega}:f:\nonumber \\
&=&\int\extd x_-\extd x_+\extd t\, e^{\imath k_-\cdot
x_-}e^{\imath k_+\cdot x_+}e^{\imath\omega t}f(x_--\lambda
k_+t,x_+,t)\nonumber\\
&=&\CF_{\rm usual}(f)(k_-,k_+,\omega+\lambda k_-\cdot
k_+),\label{thFou}\end{eqnarray} i.e. reduces to a usual Fourier
transform.

For the Laplacian, because the algebra does not have a Euclidean
covariance it is not very natural to take the usual metric and
Hodge $\star$ operator on forms, but if one does this one would
have the usual $(\del^0)^2-(\del^-)^2-(\del^+)^2$ etc. Since the
derivatives are the usual ones on normal ordered expressions, in
momentum space it becomes $\omega^2-k_-^2-k_+^2$, etc. without any
modifications. The same applies in the Maxwell theory; the only
subtlety is to keep all expressions normal ordered. On the other
hand it is not clear how logical such a metric is since there is
no relevant background symmetry of orthogonal type. Probably more
natural is a Laplacian like $(\del^0)^2-\del^-\cdot\del^+$. On the
other hand, the $U(1)$-Yang-Mills theory, as with the Chern-Simons
theory if one wants it, involves commutation relations between
functions and forms and begins to show a difference.

Finally, if we want the original (\ref{thspace}) we should
quotient by $t=1$. Then the calculus is also quotiented by $\extd
t=0$ and we see that the calculus becomes totally commutative in
the sense $[\extd x_\mu,x_\nu]=0$ as per the classical case. This
is the case relevant to the string theory literature where,
indeed, one finds that field theory etc. at an algebraic level has
just the same form as classically. This is also the conclusion in
another approach based on symmetric categories\cite{Oec:twi}.
Although a bit trivial noncommutative-geometrically, the model is
still physically interesting and we refer to the physics
literature\cite{SeiWit:non}. As well as D-branes, there are
applications to the physics of motion in background fields and the
quantum Hall effect. Moreover, the situation changes when one
considers nonAbelian gauge theory and/or questions of analysis,
where there appear instantons. From the quantum groups point of
view one should consider frame bundles and spin connections with a
frame group or quantum group of symplectic type using the
formalism of\cite{Ma:rief}. One can also use the automorphisms
provided by the quantum double of the Heisenberg algebra. These
are some topics for further development of this model.

\section{Quantum $U(1)$-Yang-Mills theory on the $\Z_2\times\Z_2$ lattice}

In this section we move to a different class of examples of
noncommutative geometry, where the algebra is that of functions on
a group lattice. It is important that the machinery is identical
to the one above, just applied to a different algebra and a
corresonding analysis of its differential calculus. In this sense
it is part of one noncommutative `universe' in a functorial sense
and not an ad-hoc construction specific to lattices. In this case
we gain meaningful answers even for a finite lattice. In usual
lattice theory this would make no sense because constructions are
justified only in the limit of zero lattice spacing, other aspects
are errors; in noncommutative geometry the finite lattice or
finite group is an exact geometry in its own right.

Specifically, we look at functions $\C[G]$ on a finite group $G$.
In this case the invariant calculi are described by ${\rm
ad}$-stable subsets $\CC\subset G$ not containing the identity.
The elements of the subset are the allowed `directions' by which
we may move by right translation from one point to another on the
group. Hence the elements of $\CC$ label the basis of invariant
1-forms $\{e_a\}$. The differentials are \eqn{Gcalc}{
\Omega^1=\C[G].\C \CC,\quad \extd
f=\sum_{a\in\CC}(\del^af)e_a,\quad \del^a=R_a-\id,\quad e_a
f=R_a(f)e_a,} where $R_a(f)=f((\ )a)$ is right translation. There
is a standard construction for the wedge product of basic forms as
well. This set up is an immediate corollary of the analysis of
\cite{Wor:dif} but has been emphasised by many authors, such as
\cite{Bre} or more recently \cite{MaRai:ele}\cite{MaSch:lat}.
Moreover, one can take a metric $\delta_{a,b^{-1}}$ in the
$\{e_a\}$ basis, which leads to a canonical Hodge $\star$
operation if the exterior algebra is finite dimensional. Then one
may proceed to Maxwell and Yang-Mills theory as well as gravity.

We now demonstrate some of these ideas in the simplest case
$G=\Z_2\times\Z_2$. It is a baby version of the treatment already
given for $S_3$ in \cite{MaRai:ele} but we take it further to the
complete quantum theory in a path integral approach. We take
$\CC=\{x=(1,0),y=(0,1)\}$ as the two allowed directions from each
point, i.e. our spacetime consists of a square with the allowed
directions being edges. The corresponding basic 1-forms are
$e_x,e_y$ and we have \[ \extd f=(\del^x f)e_x+(\del^y f)e_y,\quad
\extd e_x=\extd e_y=0,\quad e_x^2=e_y^2=\{e_x,e_y\}=0.\] The top
form is $e_x\wedge e_y$ and the Hodge $\star$ is \[ \star
1=e_x\wedge e_y,\quad \star e_x=e_y,\quad \star e_y=-e_x,\quad
\star(e_x\wedge e_y)=1.\]

We write a $U(1)$ gauge field as a 1-form $A=A^xe_x+A^y e_y$. Its
curvature \[ F=\extd A+A\wedge A=F^{xy}e_x\wedge e_y,\quad
F^{xy}=\del^x A^y-\del^y A^x+A^xR_x A^y\] is covariant as
$F\mapsto uFu^{-1}$ under \[  A\mapsto uAu^{-1}+u\extd
u^{-1},\quad A^a \mapsto {u\over R_a(u)}A^a+u\del^a u^{-1}\] for
any unitary $u$ (any function of modulus 1). We specify also the
reality condition $A^*=A$ where the basic forms are self-adjoint
in the sense $e_i^*=e_i$. This translates in terms of components
as \eqn{Areal}{ \bar A^a=R_a A^a,\quad \bar
F^{xy}=-R_{xy}(F^{xy})} under complex conjugation and implies that
$F^*=F$. Such reality should also be imposed in the examples in
the previous section if one wants to discuss Lagrangians. Finally,
we change variables by \eqn{Apolar}{ A_x+1=\lambda_x e^{\imath
\theta_x},\quad A_y+1=\lambda_y e^{\imath\theta_y},} where the
$\lambda_x,\lambda_y\ge 0$ are real and $\theta_x,\theta_y$ are
angles. The reality condition means in our case that a connection
is determined by  real numbers \eqn{lai}{
\lambda_1=\lambda_x(x),\quad \lambda_2=\lambda_y(x),\quad
\lambda_3=\lambda_x(y),\quad \lambda_4=\lambda_y(y)} and similarly
$\theta_1,\cdots,\theta_4$.

With these ingredients the Yang-Mills Lagrangian $\CL  e_x\wedge
e_y =-{1\over 2}F^*\wedge \star F$, along the same lines as for
any finite group, takes the form (discarding total derivatives),
\eqn{Z2L}{\CL= {1\over 2}|F^{xy}|^2={1\over
2}(\lambda^2_x\del^x\lambda_y^2
+\lambda^2_y\del^y\lambda^2_x)+\lambda_x^2\lambda^2_y
-\lambda_x\lambda_yR_y(\lambda_x)R_x(\lambda_y)w_1} where the
Wilson loop \eqn{wilson}{ w_1=\Re\, e^{\imath\theta_x}e^{\imath
R_x\theta_y}e^{-\imath R_y\theta_x}e^{-\imath \theta_y}} is the
real part of the holonomy around the square where we displace by
$x$, then by $y$, then back by $x$ and then back by $y$, as
explained in \cite{MaRai:ele}. When we sum over all points on the
lattice, we have the action \eqn{Z2S}{ \CS=\sum
\CL=(\lambda_1^2+\lambda_3^2)(\lambda_2^2+\lambda_4^2)
-4\lambda_1\lambda_2\lambda_3\lambda_4\cos(\theta_1-\theta_2+\theta_3-\theta_4).}

We now quantise this theory in a path integral approach. We
`factorise' the partition function into one where we first hold
the $\lambda_i$ variables fixed and do the $\theta_i$ integrals,
and then the $\lambda_i$ integrals. The first step is therefore a
lattice $U(1)$-type theory with fixed $\lambda_i$. The latter are
somewhat like a background choice of `lengths' associated to our
allowed edges. We refer to \cite{MaRai:ele} for a discussion of
the interpretation. We assume the usual measure on the gauge
fields before making the polar transformation, then \eqn{ZZla}{
\CZ=\int\extd^4\lambda\,
 \extd^4\theta\, \lambda_1\lambda_2\lambda_3\lambda_4\, e^{-\alpha
\CS}=\int_0^\infty \extd^4\lambda\,
\lambda_1\lambda_2\lambda_3\lambda_4\,
e^{-\alpha(\lambda_1^2+\lambda_3^2)(\lambda_2^2+\lambda_4^2)}\CZ_\lambda}
where $\alpha>0$ is a coupling constant and
\begin{eqnarray}\CZ_\lambda&=&\int_0^{2\pi}\extd^4\theta\,
e^{4\alpha\lambda_1\lambda_2\lambda_3\lambda_4\cos(\theta_1-\theta_2+\theta_3-\theta_4)}\nonumber
\\
&=&\int_0^{2\pi}\extd\theta_2\extd\theta_3\extd\theta_4
\int_{-\theta_2+\theta_3-\theta_4}^{2\pi-\theta_2+\theta_3-\theta_4}
\extd\theta\,
e^{\beta\cos(\theta)}\nonumber\\
&=&(2\pi)^3\int_0^{2\pi}\extd\theta\,
e^{\beta\cos(\theta)}=(2\pi)^4I_0(\beta)\label{Zla}\end{eqnarray}
where $\beta=4\alpha\lambda_1\lambda_2\lambda_3\lambda_4$ and
$I_0$ is a Bessel function, and we changed variables to
$\theta=\theta_1-\theta_2+\theta_3-\theta_4$. For the expectation
values of Wilson loops in this $U(1)$ part of the theory we take
the real part of the holonomy $w_n=\cos(n\theta)$ along a loop
that winds around our square $n$ times. Then similarly to the
above, we have \eqn{vevwilson}{
\<w_n\>_\lambda=\<\cos(n\theta)\>_\lambda={I_n(\beta)\over
I_0(\beta)}.} In other words usual Fourier transform on a circle
becomes under quantization a Bessel transform,
\[ \<{\rm Fourier\ Transform}\>_\lambda  \cdot \CZ_\lambda
={\rm Bessel\ transform}\]
\[ \<\sum_n
a_n\cos(n\theta)\>_\lambda\cdot\CZ_\lambda=(2\pi)^4\sum_n a_n
I_n(\beta)\] as the effect of taking the vacuum expectation value.
The left hand side here is the unnormalised expectation value. The
same results apply in a `Minkowski' theory where $\alpha$ is
replaced by $\imath\alpha$, with Bessel $J$ functions instead. Of
course, the appearance of Bessel functions is endemic to lattice
theory; here it is not an approximation error but a clean feature
of the $\Z_2\times\Z_2$ theory.

At the other extreme, we consider the reverse order in which the
$\theta_i$ integrals are deferred. We change variables by
\[ \lambda_1={1\over 2}\sqrt{x}(a+\sqrt{2-a^2}),\quad \lambda_2={1\over
2}\sqrt{y}(b+\sqrt{2-b^2})\]
\[ \lambda_3={1\over
2}\sqrt{x}(-a+\sqrt{2-a^2}),\quad \lambda_4={1\over
2}\sqrt{y}(-b+\sqrt{2-b^2})\] where $a,b\in[-1,1]$ and $x,y>0$.
Then
\[\lambda_1^2+\lambda_3^2=x,\quad \lambda_2^2+\lambda_4^2=y,\quad
\lambda_1\lambda_2\lambda_3\lambda_4={1\over 4}xy(1-a^2)(1-b^2)\]
\begin{eqnarray*}\CZ&=&(2\pi)^3\int_0^{2\pi}\extd\theta \CZ_\theta\nonumber\\
 \CZ_\theta&=& {1\over 16}\int_0^{\infty}\extd x\extd
y\int_0^1\extd a\extd b\, e^{-\alpha
xy(1-\cos(\theta)(1-a^2)(1-b^2))}{xy(1-a^2)(1-b^2)\over
\sqrt{(2-a^2)(2-b^2)}}.\end{eqnarray*} Only the product $z=xy$ is
relevant here and moving to this and the ratio $x/y$ as variables,
the latter gives a logarithmically divergent constant factor which
we discard, leaving the $z$-integral, which we do. We let
\[ A(a,b)= (1-a^2)(1-b^2).\] Then up to an overall factor
\begin{eqnarray} \CZ_\theta
&=& \int_0^\infty\extd z\int_0^1\extd a\extd b\,
 {Az e^{-\alpha z(1-\cos(\theta)A)}\over \sqrt{(2-a^2)(2-b^2)}}\nonumber \\
&=&{1\over\alpha^2}\int_0^1\extd a\extd b{A\over
(1-\cos(\theta)A)^2\sqrt{(2-a^2)(2-b^2)}}.\label{Zth}\end{eqnarray}
This integral is convergent for all $\theta\ne 0$. The
unnormalised expectation of the `scale Wilson loop' is similarly
\[
\<\lambda_1\lambda_2\lambda_3\lambda_4\>_\theta\cdot\CZ_\theta={1\over
2\alpha^3} \int_0^1\extd a\extd b{A^2\over
(1-\cos(\theta)A)^3\sqrt{(2-a^2)(2-b^2)}}\] which is again
convergent. If $\theta=0$ we have divergent integrals but can
regularise the theory by, for instance, doing the $a$-integrals
from $\eps>0$, giving $1/\eps^2$ and $1/\eps^4$ divergences
respectively. We plot
$\<\lambda_1\lambda_2\lambda_3\lambda_4\>_\theta$ in Figure~2 with
$\alpha=1$. There is a similar but sharper appearance to
$\CZ_\theta$ itself. The point $\theta={\pi\over 2}$ is special
and has values
\[ \CZ_{\pi\over 2}={1\over 4},\quad
\<\lambda_1\lambda_2\lambda_3\lambda_4\>_{\pi\over 2}={\pi^2\over
32},\] while the minima occur at $\theta=\pi$ with small positive
value.
\begin{figure}
\[ \epsfbox{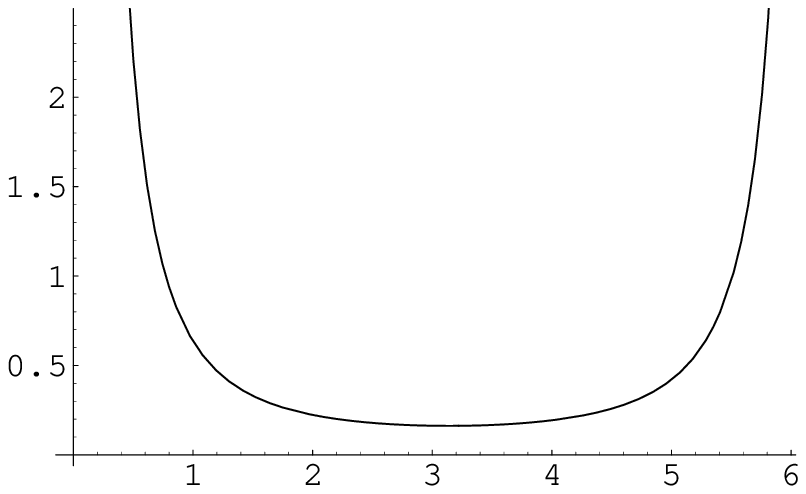}\]
\caption{Vacuum expectation values
$\<\lambda_1\lambda_2\lambda_3\lambda_4\>_\theta$ plotted for
$\theta\in [0,2\pi]$ and $\alpha=1$}
\end{figure}

Finally, we look at the full theory. From our first point of view,
for the expectation of the $U(1)$-Wilson loops in the full theory,
we need integrals
\begin{eqnarray*} && \kern-20pt \<(\lambda_1\lambda_2\lambda_3\lambda_4)^m
w_n\>\cdot\CZ\\
&&=(2\pi)^4\int_0^\infty\extd^4\lambda\,
 (\lambda_1\lambda_2\lambda_3\lambda_4)^{m+1}
  e^{-\alpha (\lambda_1^2+\lambda_3^2)(\lambda_2^2+\lambda_4^2)}
I_n(4\alpha\lambda_1\lambda_2\lambda_3\lambda_4). \end{eqnarray*}
We change the $\lambda_i$ variables to $z,a,b$ and discard a log
divergent factor as before. Then up to an overall constant,
\eqn{Zfull}{ \<(\lambda_1\lambda_2\lambda_3\lambda_4)^m
w_n\>\cdot\CZ=\int_0^{\infty}\extd z \int_0^1\extd a\extd b\,
{z^{m+1} A^{m+1} e^{-\alpha z}I_n(\alpha z A)\over 4^{m+1}
\sqrt{(2-a^2)(2-b^2)}}.} For example, we have
\begin{eqnarray*} \CZ&=&{1\over
4\alpha^2} \int_0^1\extd a\extd b\, {A\over (1-A^2)^{3\over 2}
\sqrt{(2-a^2)(2-b^2)}}
\\
\<\lambda_1\lambda_2\lambda_3\lambda_4\>\cdot\CZ&=&{1\over
16\alpha^3} \int_0^1\extd a\extd b\, {A(2+A^2)\over
(1-A^2)^{5\over 2}
\sqrt{(2-a^2)(2-b^2)}}\\
\<w_1\>\cdot\CZ &=& {1\over 4 \alpha^2} \int_0^1\extd a\extd b\,
{A^2\over (1-A^2)^{3\over 2} \sqrt{(2-a^2)(2-b^2)}}
\end{eqnarray*}
These are all divergent, with $Z\sim 1/\alpha^2\eps$ and so on.
The physical reason is the singular contribution from
configurations in the functional integral where $\theta=0$ as seen
above. Moreover, if we regularise by doing the $a$-integral (say)
from $\eps>0$ then \eqn{lambeps}{
\<(\lambda_1\lambda_2\lambda_2\lambda_4)^m\>\sim {1\over
\alpha^m\eps^{2m}}\sim
\<(\lambda_1\lambda_2\lambda_2\lambda_4)^mw_1\>.} In particular,
we have a finite answer $\<w_1\>=1$.

In summary, for the full theory we see that some of the physical
vacuum expectation values connected with the phase part of the
noncommutative gauge theory, such as the Wilson loop $\<w_1\>$,
are finite but not necessarily interesting. However, if one makes
$\alpha$ itself a divergent function of the regulator (such as
$\alpha=1/\eps^2$), one can render all the
$\<(\lambda_1\lambda_2\lambda_3\lambda_4)^m\>$ and
$\<(\lambda_1\lambda_2\lambda_3\lambda_4)^mw_1\>$ finite as well
and indeed one obtains nontrivial answers. A systematic treatment
will not be attempted here (one should choose the regulator more
physically,  among other things) but it does appear that the
theory is largely renormalisable.

\section{Clifford algebras as noncommutative spaces}

An idea for `quantisation' in physics is the Moyal product where
we work with the same vector space of functions on $\R^n$ but
modify the product to a noncommutative one. Exactly the same
multiplication-alteration idea can be used to construct Clifford
algebras on the vector space $\C[(\Z_2)^n]$ but modifying its
product. The formulae are similar but instead of real-valued
vectors $x\in\R^n$ we work with $\Z_2$-valued vectors, so that
Clifford algebras are in a precise sense discrete `quantisations'
of the $(\Z_2)^n$ lattice. We will explain this a bit more from
the Drinfeld-twist point of view in
\cite{AlbMa:qua}\cite{AlbMa:cli}, but for the present purposes the
general idea of building Clifford algebras by
multiplication-alteration factors is well known. It is mentioned
in \cite{Dix} for example, as well as in the mathematics
literature where Clifford algebras were constructed as twisted
$(\Z_2)^n$ group rings in \cite{CanOst}. On the other hand, our
point of view tells us immediately how to do the differential
calculus etc. on such spaces and that one has the same cohomology,
eigenvalues of the Laplace operator etc. as the `classical'
untwisted case. Before doing this, Section~4.1 finishes the
differential geometry for the `classical' case $(\Z_2)^n$ as in
Section~3, but now for general $n$.

\subsection{Cohomology and Maxwell theory on $(\Z_2)^n$}

We use the setting for finite groups as in Section~3, but now
functions are on $(\Z_2)^n$. For $\CC$ we take the $n$ directions
where we step $+1$ in each of the $\Z_2$ directions, leaving the
others unchanged. The differentials are \eqn{delZ2n}{ (\del^a
f)(x)=f(x+(0,\cdots,1,\cdots 0))-f(x),} where $1$ is in the $a$-th
position. We denote by $e_a$ the basic 1-forms, so that
\eqn{Z2ncalc}{ \extd f=\sum_{a=1}^n(\del^af)e_a,\quad e_a
f=R_a(f)e_a,\quad e_a^2=0,\quad \{e_a,e_b\}=0,\quad \extd e_a=0}
where $R_a(f)=f(x+(0,\cdots,1,\cdots 0))$ in the $a$-th place.
Because of the cyclic nature of the group $\Z_2$ it is easy to see
that no function can obey $\del^1f=1$ (or similarly in any other
direction). From this kind of argument one can find that the
noncommutative de Rham cohomology (the closed forms modulo the
exact ones) is represented by the grassmann algebra $\Lambda$ of
basic forms generated by the $\{e_a\}$ with the anticommutativity
relations in (\ref{Z2ncalc}), \eqn{Z2ncoh}{
H^\cdot(\C[(\Z_2)^n])=\Lambda.} Here $H^1=\C e_1\oplus\cdots\oplus
\C e_n$ has the same form as for a classical torus
$S^1\times\cdots\times S^1$ and for the same reason (and the same
holds for $(\Z_m)^n$ for all $m$). In particular, the top form is
$e_1\cdots e_n$ and is not exact. Moreover, because the exterior
algebra among the basic forms is just the usual one, the Hodge
$\star$ is the usual one given by the totally antisymmetric
epsilon tensor. We take the Euclidean metric $\delta_{ab}$ on the
basic forms. Then the spin zero wave operator is \eqn{Z2nlap}{
\square={1\over 4}\star\extd\star\extd={1\over
4}\sum_a\del^a\del^a=-{1\over 2}\sum_a \del^a.}

The plane wave eigenfunctions of the partial derivatives $\del^a$
are \eqn{Z2nwave}{ \psi_k(x)=(-1)^{k\cdot x},\quad
\del^a\psi_k=-2k_a\psi_k} labelled by momenta $k\in(\Z_2)^n$.
These diagonalise the wave operator, which has eigenvalue
\eqn{Z2nlapeval}{ \square\psi_k=|k|\psi_k,\quad |k|=\sum_a
k_a=k\cdot k} since each $k_a\in\{0,1\}$. The eigenvalues range
from $0,\cdots,n$ and there are $\left({n\atop m}\right)$
eigenfunctions for eigenvalue $m$. In particular, there are no
massless modes other than the constant function $1$.

For the spin 1 equation, since the formulae among basic 1-forms
are as for $\R^n$, the Maxwell operator $\star\extd\star\extd$ on
1-forms $A=\sum_a A^a e_a$ is given by the above $\square$ on each
component function if $A$ is in Lorentz gauge $\del\cdot A=0$. Its
eigenvalues therefore also range from $0,\cdots,n$ according to
the degree of all components. The Maxwell operator on the
$(n-1)2^n+1$ dimensional space of gauge fields $A$ such that
$\del\cdot A=0$ can be fully diagnonalised by the following modes.
There are (i) $n2^{n-1}$ modes of the form \eqn{Z2nmaxi}{
A=\psi_{k}e_a} of degree $|k|=0,\cdots,n-1$, where $a=1,\cdots,n$
and $k$ is such that its $a$-component $k_a$ vanishes. In addition
there are (ii) $\left({n\atop m}\right)(m-1)$ modes of degree
$m=2,\cdots,n$ of the form \eqn{Z2nmaxii}{ A=\psi_{k}(\mu_1
e_{a_1}+\cdots+\mu_m e_{a_m}),\quad \mu_1+\cdots\mu_m=0,} where
the momentum vector $k$ has components $k_{a_1}=\cdots=k_{a_m}=1$
and the rest zero. Adding up the modes of type (ii) gives
$n2^{n-1}+1-2^n$, which with the modes of type (i) gives a total
of $(n-1)2^n+1$, which is the required number since the space of
$A$ is $n2^n$-dimensional while the image of $\extd$, i.e. the
number of exact forms, is $2^n-1$-dimensional. Note that the gauge
fixing by $\del\cdot A$ still leaves possible $A\mapsto A+\extd f$
where $\square f=0$ but then $f=1$ by the above, and $\extd f=0$,
i.e. the gauge-fixing is fully effective. These eigenfunctions $A$
with degree $>0$ are also the allowed sources, being a basis of
solutions of $\del\cdot J=0$ in the image of the Maxwell wave
operator. Hence we can solve fully the Maxwell equations. One
could also rework these results with a Minkowski metric in the
$e_a$ basis.

\subsection{Noncommutative geometry on Clifford algebras}

In this section we look at the standard Clifford algebra ${\rm
Cliff}(n)$  with generators $\gamma_a$, $a=1,\cdots,n$ and
relations \eqn{cliff}{ \{\gamma_a,\gamma_b\}=\delta_{a,b}.} In
\cite{CanOst,AlbMa:cli} this is presented as a twist of the group
algebra of $(\Z_2)^n$. This is the momentum group for our
classical space $\C[(\Z_2)^n]$ above, with elements the plane
waves $\psi_{k}$. The `quantisation' procedure consists of
changing the product between such basis vectors to a new one
\eqn{psiprod}{ \psi_k\bullet\psi_m=F(k,m)\psi_{k+m},\quad
F(k,m)=(-1)^{\sum_{j<i} k_i m_j}} where $F$ is a cocycle on the
group $(\Z_2)^n$ just because it is bilinear. These relations are
such that if we write\eqn{psia}{
\psi_a=\psi_{(0,\cdots,1,\cdots,0)},\quad
\psi_a(x)=\begin{cases}-1& {\rm if}\ x_a=1\\ 0& {\rm
else}\end{cases},} where $1$ in the momentum vector is in the
$a$-th position, then
\[ \psi_a\bullet\psi_b+\psi_b\bullet\psi_a=\delta_{a,b}\]
so that we can identify $\gamma_a=\psi_a$. We similarly define
\eqn{gammax}{ \gamma_k=\gamma_1^{k_1}\cdots\gamma_n^{k_n}} for any
$k\in(\Z_2)^n$ and identify $\gamma_k=\psi_k$. In this way the
whole basis of $\C[(\Z_2)^n]$ is identified with the vector space
of the Clifford algebra. This approach makes ${\rm Cliff}(n)$ a
braided-commutative algebra in a symmetric monoidal category, see
\cite{AlbMa:cli}.

We can similarly identify all normal ordered expressions where the
$\gamma_a$ occur in increasing order from left to right with the
identical `classical' expression in terms of the commutative
$\psi_a$. Moreover, because the Clifford algebra is obtained by a
cocycle twist, it follows from the general theory in the next
section that we may likewise identify the differential calculi.
Indeed, the entire exterior algebra $\Omega({\rm Cliff}(n))$ is a
twist by the same cocycle $F$ but now of the exterior algebra of
differential forms on $\C[(Z_2)^n]$ from Section~4.1, with the
same cohomology and gauge theory. Explicitly, \eqn{extcliff}{
\Omega({\rm Cliff}(n))={\rm Cliff}(n).\Lambda,} where $\Lambda$ is
our previous grassmann algebra (\ref{Z2ncalc}) generated by the
anticommuting $e_a$. We have \eqn{extdcliff}{\extd f=\sum_a
(\del^a f)e_a,\quad \del^a:f:=:\del^a f:,} where $:f:$ is the
normal ordered element defined by writing $f$ in terms of the
$\psi_a$ and on the right we use the partial derivatives from the
previous section. Similarly for the noncommutation relations
between functions and 1-forms. Explicitly, \eqn{cliffcalc}{
\del^a\gamma_k=-2k_a\gamma_k,\quad
e_a\gamma_b=\begin{cases}-\gamma_b e_a&{\rm if}\ a=b\\ \gamma_b
e_a&{\rm else}\end{cases}.}

Taking the $\delta_{ab}$ metric on the $e_a$ basis as before, and
the corresponding Hodge $\star$ etc., we have the Laplace operator
${1\over 4}\star\extd\star\extd$ on plane waves \eqn{clifflap}{
\square \gamma_k=|k|\gamma_k} so that the total degree function on
the Clifford algebra has a nice geometrical interpretation as the
eigenvalue of the Laplacian. Similarly, the diagonalisation of the
wave operator in spin 1 has the same form as in Section~4.1. For
example, in ${\rm Cliff}(2)$ there are five eigenfunctions with
$\del\cdot A=0$, namely
\[ e_1,\ e_2,\ \gamma_2e_1,\ \gamma_1e_2,\
\gamma_1\gamma_2(e_1-e_2).\] The allowed sources in the Maxwell
theory are spanned by the latter three. The corresponding
solutions for $A$ and their curvatures are respectively
\[ A=\gamma_2 e_1,\ \gamma_1 e_2,\ {1\over
2}\gamma_1\gamma_2(e_1-e_2),\quad F=\extd A=2\gamma_2 e_1\wedge
e_2,\ -2\gamma_1 e_1\wedge e_2,\  2\gamma_1\gamma_2e_1\wedge
e_2.\]

One also has a $U(1)$-Yang-Mills theory on the Clifford algebra,
where the curvature is $\extd A+A\wedge A$ and gauge transform is
(\ref{Agauge}) under an invertible element of the Clifford
algebra. The subgroup of such gauge transformations restricted to
preserve the generating space $\{\gamma_a\}$ under the Clifford
adjoint operation is the Clifford group, giving a geometrical
interpretation of that.

\section{Twisting of differentials and the noncommutative torus
$A_\theta$.}

In this more technical section we explain the theorem for the
systematic twisting of differentials on quantum groups and a small
corollary of it which covers our above treatment of differential
calculus on Clifford algebras as well as on the noncommutative
torus and other spaces. The theory works at the level of a Hopf
algebra $H$ but for the purposes of the present article this could
just be the coordinate algebra $\C[G]$ of an algebraic group or
the group algebra of a discrete group; Hopf algebras are a nice
way to treat the continuum and finite theories on an identical
footing, as well as allowing one to generalise to quantum groups
if one wants $q$-deformations etc.

A 2-cocycle on the group in our terms is an element $F:H\tens H\to
\C$ obeying the condition \eqn{cocycleH}{ F(b\o\tens c\o)F(a\tens
b\t c\t)=F(a\o\tens b\o)F(a\t b\t\tens c),\quad F(1\tens
a)=\eps(a)}for all $a,b,c\in H$, where $\Delta a=a\o\tens a\t\in
H\tens H$ is our notation for the coproduct. We will similarly
denote $(\id\tens \Delta)\Delta a=(\Delta\tens\id)\Delta
a=a\o\tens a\t\tens a\th$ for the iterated coproduct. On a group
manifold, the counit $\eps$ simply evaluates at the group
identity. It is a construction going back to Drinfeld that in this
case there is a new Hopf algebra $H_F$ with product \eqn{twistH}{
a\bullet b=F(a\o\tens b\o) a\t b\t F^{-1}(a\th\tens
b\th),\quad\forall a,b\in H} where $F^{-1}$ is the assumed
convolution inverse. See \cite{Ma:book} for details and proofs.
Another fact is that for a bicovariant calculus on a Hopf algebra
$H$, the exterior algebra $\Omega(H)$ is a super-Hopf
algebra\cite{Brz:rem}. Then:

\begin{theorem} (S.M. \& R. Oeckl \cite{MaOec:twi}) Let $F$ be
a 2-cocycle on a Hopf algebra $H$ with bicovariant calculus
$\Omega(H)$. Extend $F$ to a super-cocycle on $\Omega(H)$ by zero
on degree $>0$. Then $\Omega(H_F)=\Omega(H)_F$ is a bicovariant
calculus on $H_F$.
\end{theorem}

This took care of how calculi on Hopf algebras themselves behave
under twisting. We need a slight but immediate extension of this.
Let $A$ be an algebra on which $H$ coacts by an algebra
homomorphism $\Delta_L:A\to H\tens A$. We will use the notation
$\Delta_L a=a\bo\tens a\bt$. In the case when $H$ is functions on
a group, given any group element we evaluate the $H$ part of the
output of $\Delta_L$ on the element and have a map $A\to A$, i.e.
a coaction is just the same thing as a group action but expressed
in terms of the coordinates on the group. A fundamental theorem
\cite{Ma:euc} is that when we twist $H$ to $H_F$ we must also
twist $A$ to a new algebra $A_F$ for it to be $H_F$-covariant; its
product is \eqn{twistA}{ a\bullet b=F(a\bo\tens b\bo) a\bt
b\bt,\quad \forall a,b\in A.} There is obviously the same theorem
in the category of super algebras coacted upon by super-Hopf
algebras, which is trivial to spell out (one just puts the signed
super-transposition in place of the usual transposition in all
constructions). Now suppose that $A$ is covariant under $H$ and is
equipped with a covariant differential calculus such that the
coaction extends to all of the exterior super-algebra $\Omega(A)$
of (possibly noncommutative) differential forms as a comodule
algebra under $H$. So we can apply (\ref{twistA}) to $\Omega(A)$
in place of $A$. Clearly,

\begin{corol} If $\Omega(A)$ is $H$-covariant on an
$H$-covariant algebra $A$ and if $F$ is a cocycle on $H$ then
$\Omega(A_F)=\Omega(A)_F$ is an $H_F$-covariant calculus on $A_F$.
\end{corol} One could also view $\Omega^1(A)$ as a
$\Omega^1(H)$-super comodule algebra and then using Theorem~5.1
and the super-version of (\ref{twistA}) applied to $\Omega(A)$ we
see in fact that $\Omega(A_F)$ is an $\Omega(H_F)$-super comodule
algebra.

Since any Hopf algebra coacts covariantly on itself via
$\Delta_L=\Delta$, a canonical example is to take $A=H$ with this
coaction. Then for any cocycle $F$ we have a new algebra $A_F$ via
(\ref{twistA}) covariant under $H_F$ obtained via (\ref{twistH}).
This was the theory behind \cite{AlbMa:qua,AlbMa:cli}. Now suppose
that $\Omega(H)$ is a bicovariant calculus on $H$, then it is in
particular left covariant. So $\Omega(A)=\Omega(H)$ is left
$H$-covariant as induced by $\Delta_L$ on functions. We can
therefore apply Corollary~5.2 to obtain $\Omega(A_F)$ also. In
fact\cite{Wor:dif} $\Omega(A)\isom A\tens \Lambda$ where $\Lambda$
is the algebra of left-invariant forms generated by $\{e_a\}$ say.
The coaction on these forms is trivial and hence when we apply
(\ref{twistA}) to $\Omega(A)$ we find simply
\begin{eqnarray} e_a\wedge_\bullet e_b&=&F(e_a\bo\tens
e_b\bo)e_a\bt\wedge e_b\bt=e_a\wedge e_b\nonumber \\
 a\bullet e_a&=&F(a\bo\tens e_a\bo)a\bt
e_a\bt=ae_a\nonumber \\
 e_a\bullet a&=&F(e_a\bo\tens a\bo)e_a\bt a\bt=e_a a,\quad \forall a\in
A.\label{twistbimod}\end{eqnarray} So the form relations are
unchanged in this basis $\{e_a\}$. We get the same result in the
super-twisting point of view: the supercoproduct in $\Omega(H)$
looks like $\und \Delta e_a = \Delta_Re_a +1\tens e_a$ where
$\Delta_R e_a\subset \Lambda^1\tens H$ is a certain right
coaction. Here $\Lambda^1$ is spanned by the $\{e_a\}$. Since the
cocycle is trivial on 1 in the sense in (\ref{cocycleH}) and
similarly for $F(a\tens 1)$, and since extended by zero on the
$e_a$, we see that the super version of ({\ref{twistA}) with
$\Omega^1(A)=\Omega^1(H)$ and supercoaction
$\und\Delta_L=\und\Delta$ gives the same result that all of these
products involving 1-forms are unchanged.

Let us see how all of this works for a discrete group $G$. Here
$H=\C G$ is the group algebra spanned by group elements, with
coproduct $\Delta g=g\tens g$ for all $g\in G$. The equation
(\ref{cocycleH}) then reduces to a group cocycle. Yet, because $H$
is cocommutative, the twisting (\ref{twistH}) has no effect and
$H_F=H$ is unchanged for all $F$. On the other hand, we can take
$A=\C G$ as a left-covariant algebra under $\Delta_L=\Delta$. This
time (\ref{twistA}) means a new algebra $A_F$ with product
\eqn{twistG}{ g\bullet h=F(g,h) gh,\quad \forall g,h\in G,}
extended linearly. This is the special case used in Section~4.

We now look at the initial differential structure. To keep the
picture simple we assume $G$ is Abelian (this is not necessary).
Then under Fourier transform $\C G\isom \C[\hat G]$, where $\hat
G$ is the group of characters (if $G$ is infinite then this will
be compact and $\C[\hat G]$ is the algebraic coordinate ring). In
the compact case we use indeed the classical calculus on this
`position space' $\hat G$ while in the finite case we use the
set-up (\ref{Gcalc}) for a chosen conjugacy class. That is the
geometrical picture, but we do not have to actually do the Fourier
transform to the coordinate ring picture, we rather work directly
with the `momentum group' $G$ dual to the position space. Then
\eqn{OmegaCG}{ \Omega(\C G)=(\C G)\Lambda} where $\Lambda$ here is
the usual grassmann algebra of basic forms $\{e_a\}$. The
differential and relations have the form \eqn{hatGcalc}{ \extd
g=\sum_a (\chi_a(g)-1)g e_a,\quad e_a g=\chi_a(g)ge_a,\quad
\forall g\in G} and the super-coproduct is \eqn{supercopG}{
\und\Delta g=g\tens g,\quad \und\Delta e_a=e_a\tens 1+ 1\tens
e_a.} Here $\{\chi_a\}$ are some subset of characters, the allowed
directions in $\hat G$ that define the calculus. This is
equivalent to our treatment in Section~4.1 for $G=(\Z_2)^n$, with
$g$ in the role of plane waves $\psi_k$.

Next we take $A=\C G=H$ another copy of the same algebra, $F$ a
group cocycle and $\Delta_L=\Delta$  to get our twisted version
$A_F=(\C G)_F$ as in (\ref{twistG}). And we take $\Omega(A)=(\C
G)\Lambda=\Omega(H)$ as above. Then from (\ref{twistbimod}) we
know that $\Omega((\C G)_F)$ has the identical form to the
untwisted case. The only part that changes is the algebra $(\C
G)_F$ itself which becomes twisted and typically (depending on
$F$) noncommutative. This is the reason for exactly the same form
of calculus on ${\rm Cliff}(n)$ in Section~4.2 as for $(\Z_2)^n$.

Now let us cover the celebrated noncommutative torus $A_\theta$ in
the same way. This has the relations $vu=e^{\imath\theta}uv$ where
$\theta$ is an angle. For our initial situation we take
$G=\Z\times \Z$ in (\ref{hatGcalc}), with free commuting
generators $u,v$. Here the character group is $S^1\times S^1$ and
the calculus is defined by two characters
\[ \chi_{1,\phi}(u^m v^n)=e^{\imath \phi m},\quad
\chi_{2,\phi}(u^m v^n)=e^{\imath \phi n}\] with $\phi$ as a
parameter. The only difference is that we rescale $\extd$ and then
take the limit $\phi\to 0$ of this family of calculi, giving
\begin{eqnarray*} \extd (u^m v^n)&=&\lim_{\phi\to 0} {1\over
\imath\phi}\left((\chi_{1,\phi}(u^m v^n)-1)u^m v^n
e_1+(\chi_{2,\phi}(u^m v^n)-1)u^m v^n e_2\right)\\
&=&u^m v^n(me_1+n e_2)\\ e_a u^m v^n&=&u^m v^n e_a.\end{eqnarray*}
This is nothing but the usual classical differential calculus on
$\C[S^1\times S^1]$, but written algebraically and in momentum
space. Next, on $\Z\times\Z$ we take cocycle \eqn{Ftheta}{ F(u^m
v^n,u^s v^t)=e^{\imath \theta ns},} where $\theta$ denotes a fixed
parameter. Then the formula (\ref{twistG}) gives
\eqn{Atheta}{v\bullet
u=e^{\imath\theta}uv=e^{\imath\theta}u\bullet v} which is
noncommutative torus from our algebraic point of view. That its
product is a twisting is known to experts from another point of
view \cite{Rie:twi} and also for more general
$\theta$-spaces\cite{ConLan} from the twisting point of view in
recent work \cite{Sit:twi,Var:sym}. That one automatically gets
the differential geometry as a twist seems to be less well-known.
Thus, for the differential structure we obtain \eqn{torcalc}{
\extd u=u e_1,\quad \extd v=v e_2,\quad e_a u=ue_a,\quad e_a v=v
e_a} since this is the same as the classical form. Likewise the
wedge products are as per the classical form in which the $e_a$
anticommute. Note then that
\[ (\extd v)u=v e_1 u= v\bullet u e_1= e^{\imath\theta} u \bullet
v e_1=e^{\imath\theta} u \extd v\] \[ \extd v\wedge \extd
u=ve_2\wedge ue_1=e_2\wedge v\bullet u
e_1=e^{\imath\theta}e_2\wedge u\bullet v
e_1=-e^{\imath\theta}\extd u\wedge \extd v\] which is how the
calculus is usually presented, as noncommutation between $\extd
u,v$ etc., but we see that it expresses nothing more than the
noncommutativity of $A_\theta$ itself. This goes some way towards
explaining why one can develop Yang-Mills and other geometry on a
noncommutative torus\cite{ConRie} so much like on a usual torus,
with two commuting derivations as vector fields, etc. For a
general noncommutative algebra one does not expect many
derivations.

We can just as easily apply this theory to $H=A=\C[\R^n]$. Here
$H$  has the linear coproduct $\Delta x_\mu=x_\mu\tens 1+1\tens
x_\mu$ and on it we take the cocycle \eqn{moycocy}{F(f\tens
g)=e^{\imath\sum \theta_{\mu\nu}\del^\mu\tens \del^\nu}(f\tens
g)(0)} where we apply differential operators on functions $f,g$
and evaluate at zero. Then the algebra $\C[\R^n]_F$ has product
\eqn{moyprod}{ f\bullet g=\left(\sum_{m=0}^\infty {(\sum
\theta_{\mu\nu}\del^\mu\tens \del^\nu)^m\over m!}|_0(f\o\tens
g\o)\right) f\t g\t=\cdot
e^{\imath\sum_{\mu,\nu}\theta_{\mu\nu}\del^\mu\tens
\del^\nu}(f\tens g)} where $(\del^m|_0 f\o) f\t=\del^m f$. This is
the Moyal product and this point of view was explored in
\cite{MaOec:twi} in the context of generalising the cocycle to a
nonAbelian group rather than $\R^n$. It was picked up and
discussed explicitly in \cite{Oec:twi} as well as by other
authors. In particular, the coordinate functions $x_\mu$ with the
$\bullet$ product obey the algebra (\ref{thspace}) studied in
Section~2.3. From this point of view the bilinear form that
defines the Clifford algebra twisting (\ref{psiprod}) is like a
finite difference matrix. Meanwhile the cochain that similarly
twists $(\Z_2)^n$ to the octonions \cite{AlbMa:qua} has an
additional cubic term that is responsible for their
nonassociativity, i.e. more like a discrete `Chern-Simons' theory
as remarked there. Note that when $F$ is only a cochain our
Corollary~5.2 induces not a usual associative exterior algebra on
the octonions but a superquasialgebra in the same sense as the
octonions are a quasialgebra (i.e. associative when viewed in a
monoidal category).

Further afield, other $\theta$-manifolds have been of interest of
late, see \cite{ConDub} and references therein. It is clear that
these too are twistings as noncommutative algebras and therefore
that one can recover the algebraic side of their noncommutative
geometry by twisting the classical geometry as above. It is
assumed that $M$ is a classical manifold admitting a group action
by a compact Lie group $K\supseteq S^1\times S^1$. We assume an
algebraic description exists (otherwise one needs to do some
analysis), so that we have classical coordinate Hopf algebras
$\C[K]\to \C[S^1\times S^1]$ and a coaction $\C[M]\to
\C[K]\tens\C[M]$. We suppose that this extends to the classical
exterior algebra of $M$ also. The action of the subgroup here
corresponds to a coaction $\C[M]\to \C [S^1\times S^1]\tens
\C[M]$. Then with the same cocycle (\ref{Ftheta}), we obtain a new
algebra $\C[M]_F$ by the comodule algebra twisting theorem
(\ref{twistA}), and  we obtain a differential calculus
$\Omega(\C[M]_F)$ on it by Corollary~5.2, and ultimately an entire
twisted noncommutative geometry with a parameter $\theta$.
Equivalently, we can obviously pull back $F$ as a cocycle
$F:\C[K]\tens \C[K]\to \C$ and do all of the above with the
$\C[K]$-coaction directly (the result is the same). On the other
hand, whereas $\C[S^1\times S^1]$ does not itself twist as a Hopf
algebra, now (\ref{twistH}) gives a new Hopf algebra $\C[K]_F$ and
this coacts on $\Omega(\C[M]_F)$ by Corollary~5.2. By Theorem~5.1
we also have $\Omega(\C[K]_F)$ and this supercoacts. It seems that
these elementary deductions fit with observations from a different
point of view (not via the twisting theory) in \cite{ConDub}. We
see only an `easy' algebraic part of that theory but it does seem
to indicate some useful convergence between that operator theory
approach and the quantum groups one.

\baselineskip 14pt

\begin{thebibliography}{10}

\bibitem{Ma:ista}
S. Majid. \newblock Duality principle and braided geometry.
\newblock Springer Lec. Notes in Phys. 447:125--144, 1995.

\bibitem{Con}
A.~Connes.
\newblock {\em Noncommutative Geometry}.
\newblock Academic Press, 1994.

\bibitem{MaOec:twi}
S.~Majid and R.~Oeckl.
\newblock Twisting of quantum differentials and the {P}lanck scale {H}opf
  algebra.
\newblock {\em Commun. Math. Phys.}, 205:617--655, 1999.



\bibitem{Ma:pla}
S.~Majid.
\newblock {H}opf algebras for physics at the {P}lanck scale.
\newblock {\em J. Classical and Quantum Gravity}, 5:1587--1606, 1988.

\bibitem{Wor:dif}
S.L. Woronowicz.
\newblock Differential calculus on compact matrix pseudogroups (quantum
  groups).
\newblock {\em Commun. Math. Phys.}, 122:125--170, 1989.

\bibitem{BrzMa:gau}
T.~Brzezi\'nski and S.~Majid.
\newblock Quantum group gauge theory on quantum spaces.
\newblock {\em Commun. Math. Phys.}, 157:591--638, 1993.
\newblock Erratum 167:235, 1995.

\bibitem{Ma:rief}
S.~Majid.
\newblock Riemannian geometry of quantum groups and finite groups with
  nonuniversal differentials.
\newblock {\em Commun. Math. Phys.}, 225:131-170, 2002.

\bibitem{NML:rie}
F. Ngakeu, S. Majid and D. Lambert. \newblock Noncommutative
Riemannian geometry of the alternating group $A_4$.
\newblock {\em J. Geom. Phys.}, 42:259--282, 2002.

\bibitem{Ma:ric}
S.~Majid.
\newblock Ricci tensor and Dirac operator on {$\C_q[SL_2]$} at roots of unity.
\newblock {\em Lett. Math. Phys.}, 63:39--54, 2003.

\bibitem{Ma:reg}
\newblock S. Majid.
\newblock On q-Regularization. {\em Int. J. Mod. Phys. A},
5:4689-4696, 1990.

\bibitem{MaRue:bic}
S.~Majid and H.~Ruegg.
\newblock Bicrossproduct structure of the {$\kappa$}-{P}oincar{\'e} group and
  non-commutative geometry.
\newblock {\em Phys. Lett. B}, 334:348--354, 1994.

\bibitem{LNRT:def}
J. Lukierski, A. Nowicki, H. Ruegg and V.N. Tolstoy. \newblock ,
{$q$}-{D}eformation of {P}oincar{\'e} algebra. \newblock {Phys.
Lett. B}, 271:321, 1991.

\bibitem{AmeMa:wav}
G.~Amelino-Camelia and S.~Majid.
\newblock Waves on noncommutative spacetime and gamma-ray bursts.
\newblock {\em Int. J. Mod. Phys. A}, 15:4301--4323, 2000.

\bibitem{Ame:nat}
G. Amelino-Camelia. {\em Gravity-wave interferometers as
quantum-gravity detectors}. {\em Nature} 398:216, 1999.

\bibitem{BatMa:ang}
E.~Batista and S.~Majid.
\newblock Noncommutative geometry of angular momentum space
{$U(su_2)$}.
\newblock {\em J. Math. Phys.}, 44:107-137, 2003.

\bibitem{Sch:com}
B.J. Schroer. \newblock Combinatorial quantization of Euclidean
gravity in three dimensions, {\em preprint} math.QA/0006228.

\bibitem{Sny}
S. Snyder.
\newblock Quantized space-time. {\em Phys. Rev.}, 71:38-41, 1947.

\bibitem{Oec:twi}
R. Oeckl. \newblock Untwisting noncommutative $\R^d$ and the
equivalence of quantum field theories. \newblock {\em Nucl.Phys.
B} 581:559-574, 2000.

\bibitem{SeiWit:non}
N. Seiberg and E. Witten.
\newblock String theory and noncommutative geometry.
\newblock {\em J. High En. Phys.}, 9909:032, 1999.


\bibitem{Bre}
K. Bresser, F. Mueller-Hoissen, A. Dimakis, A. Sitarz.
\newblock
Noncommutative geometry of finite groups. {\em J. Phys. A}
29:2705--2736, 1996.


\bibitem{MaRai:ele}
S.~Majid and E.~Raineri.
\newblock Electromagnetism and gauge theory on the permutation group $S_3$.
\newblock {\em J. Geom. Phys.}, 44:129--155, 2002.


\bibitem{MaSch:lat}
S.~Majid and T.~Schucker.
\newblock $\Z_2 \times \Z_2$ lattice as a Connes-Lott-quantum group model.
\newblock {\em J. Geom. Phys.}, 43:1--26, 2002.

\bibitem{AlbMa:qua}
H. Albuquerque and S. Majid. \newblock Quasialgebra structure of
the Octonions.
\newblock {\em J. Algebra}, 220:188--224, 1999.

\bibitem{AlbMa:cli}
H. Albuquerque and S. Majid. \newblock Clifford algebras obtained
by twisting of group algebras. \newblock {\em J. Pure Applied
Algebra} 171:133-148, 2002.

\bibitem{Dix}
G.~Dixon.
\newblock {\em Division Algebras: Octonions, Quaternions,
Complex Numbers and the Algebraic Design of Physics}.
\newblock Kluwer, 1994.

\bibitem{CanOst}
S. Caenepeel, F. Van Oystaeyen. \newblock  A note on generalized
Clifford algebras and representations. \newblock {\em Comm. in
Algebra} 17:93-102, 1989.

\bibitem{Ma:book}
S.~Majid.
\newblock {\em Foundations of Quantum Group Theory}.
\newblock Cambridge Univeristy Press, 1995.

\bibitem{Ma:euc}
S. Majid. \newblock {$q$}-{E}uclidean space and quantum {W}ick
rotation by twisting. \newblock {\em J. Math. Phys.},
35:5025--5034, 1994.

\bibitem{Brz:rem}
T.~Brzezi\'nski.
\newblock Remarks on bicovariant differential calculi and exterior {{H}opf}
  algebras.
\newblock {\em Lett. Math. Phys.}, 27:287, 1993.

\bibitem{ConRie}
A. Connes and M.A. Rieffel.
\newblock Yang-Mills for noncommutative two-tori.
\newblock {\em AMS Contemp. Math. Series} 62:237--266, 1987.

\bibitem{Rie:twi}
M. Rieffel. \newblock Deformation quantization for actions of
$\R^d$.
\newblock {\em Memoirs AMS}, 106, 1993.

\bibitem{ConLan}
A. Connes and G. Landi.
\newblock Noncommutative manifolds, the instanton algebra and
isospectral deformations.
\newblock Commun. Math. Phys. 221:141-159, 2001.


\bibitem{Sit:twi}
A. Sitarz.
\newblock Twist and spectral triples for isospectral deformations.
\newblock {\em Lett. Math. Phys.} 58: 69-79, 2001.

\bibitem{Var:sym}
J. Varilly.
\newblock Quantum symmetry groups of noncommutative spheres.
\newblock {\em Commun. Math. Phys.} 221:511--523, 2001.

\bibitem{ConDub}
A. Connes and M. Dubois-Violette.
\newblock Noncommutative finite-dimensional manifolds, I:
Spherical manifolds and related examples.
\newblock {\em Commun. Math. Phys.} 230: 539-579, 2002.




\end{thebibliography}

\end{document}